# Kinetic Control of Morphology and Composition in Ge/GeSn Core/Shell Nanowires

Simone Assali,[1,2,¥,*] Roberto Bergamaschini,[3,¥,*], Emilio Scalise[3,¥,*], Marcel A. Verheijen,[4] Marco Albani[3], Alain Dijkstra,[1] Ang Li,[1,5] Sebastian Koelling,[1] Erik P.A.M. Bakkers,[1,6] Francesco Montalenti,[3] and Leo Miglio[3]

[1] Department of Applied Physics, Eindhoven University of Technology, 5600 MB Eindhoven, The Netherlands
[2] Department of Engineering Physics, École Polytechnique de Montréal, C. P. 6079, Succ. Centre-Ville, Montréal, Québec H3C 3A7, Canada
[3] L-NESS and Dept. of Materials Science, University of Milano Bicocca, 20125, Milano, Italy
[4] Eurofins Materials Science BV, High Tech Campus 11, 5656AE Eindhoven, The Netherlands
[5] Beijing University of Technology, Pingleyuan 100, 100124, P. R. China
[6] Kavli Institute of Nanoscience, Delft University of Technology, 2600 GA Delft, The Netherlands

## ABSTRACT

The growth of Sn-rich group-IV semiconductors at the nanoscale can enrich the understanding of the fundamental properties of metastable GeSn alloys. Here, we demonstrate the effect of the growth conditions on the morphology and composition of Ge/GeSn core/shell nanowires by correlating the experimental observations with a theoretical interpretation based on a multi-scale approach. We show that the cross-sectional morphology of Ge/GeSn core/shell nanowires changes from hexagonal to dodecagonal upon increasing the supply of the Sn precursor. This transformation strongly influences the Sn distribution as a higher Sn content is measured under the {112} growth front. *Ab-initio* DFT calculations provide an atomic-scale explanation by showing that Sn incorporation is favored at the {112} surfaces, where the Ge bonds are tensile-strained. A phase-field continuum model was developed to reproduce the morphological transformation and the Sn distribution within the wire, shedding light on the complex growth mechanism and unveiling the relation between segregation and faceting. The tunability of the photoluminescence emission with the change in composition and morphology of the GeSn shell highlights the potential of the **core/shell nanowire system for opto-electronic devices operating at mid-infrared wavelengths.**

Direct band gap Germanium-Tin (GeSn) alloys are at the forefront in the development of opto-electronic devices operating at mid-infrared wavelengths and are fabricated on a Silicon platform.[1] The epitaxial growth of GeSn layers is commonly performed on a Ge/Si virtual substrate (Ge-VS) and it provides full integration with the current Si-technology manufacturing processes.[2–5] However, as the lattice mismatch between GeSn and Ge increases with Sn content, the compressive strain in the GeSn layer shifts upward the compositional threshold for achieving a direct band gap in Sn-rich GeSn semiconductors.[6] In addition, the residual strain in the growing GeSn layer, after plastic relaxation, reduces the incorporation of Sn, eventually leading to segregation and phase separation, with the formation of Sn droplets at the surface.[4,7,8] This effect is especially pronounced when GeSn is grown directly on a Si substrate, where the lattice mismatch can reach values above 4 %.[9] Enhanced strain relaxation can be achieved in a one-dimensional geometry, using semiconductor nanowires (NWs) grown from a metal catalyst by the vapor-liquid-solid (VLS) method. This configuration was recently exploited to fabricate axial and radial GeSn-based NW heterostructures, reaching incorporation of Sn well above the 9 at.% threshold required to achieve a (strain-free) direct band semiconductor.[6,10–13] The use of pure Sn catalysts resulted in the growth of GeSn NWs with compositions up to 19 at.% and optical emission at mid-infrared wavelengths.[14] In addition, by controlling the segregation of Sn droplets on the NW sidewall, the growth of Sn-seeded $Ge_{0.92}Sn_{0.08}$ branches on $Ge_{0.96}Sn_{0.04}$ trunks was demonstrated.[15] When moving to a Ge/GeSn core/shell NW geometry, the lattice-mismatch is partially accommodated by transferring some strain into the Ge core, thus promoting strain relaxation in the GeSn shell without the nucleation of extended defects.[12,16,17] When the NW core diameter is smaller than the shell thickness, the core behaves as a compliant substrate, accommodating the lattice mismatch of the system without bending.[18,19] Moreover, the incorporation of Sn in Ge increases as the compressive strain is progressively relieved with increasing shell thickness, by a mechanism of compositional pulling during the shell growth.[16] However, at larger core diameters the shell will experience higher strain, eventually inducing plastic deformation with the nucleation of defects and surface roughening.[17]

In this work, we show how the morphology and composition of a GeSn shell grown on a Ge core NW are strongly dependent on the growth conditions. At a higher supply of the Tin-tetrachloride ($SnCl_4$) precursor, the symmetry of the NW cross-section changes by the evolution of the corners in between the {112} facets into {110} facets. At the same time, enhanced segregation is observed, with an increasing difference in composition between Sn-poor <110>-oriented stripes and Sn-rich {112} facets. In addition, at the highest supply of the Sn precursor, phase separation occurs and multiple Sn droplets are visible on the NW sidewall. The experimental observations are then rationalized theoretically by a multi-scale approach. First, the shape transition will be interpreted by a continuum kinetic growth model, including surface diffusion. Then, first principle calculations will be exploited to assess the origin of the different compositions within the facets and to extend the growth model in order to simultaneously trace the evolution of shape and composition. The agreement between experiments and theory highlights the strong correlation between faceting and segregation dynamics in the Ge/GeSn core/shell NW system.

## RESULTS AND DISCUSSION

**Epitaxial growth.** Arrays of <111>-oriented Ge/GeSn core/shell NWs were grown on a Ge (111) wafer in a chemical vapor deposition (CVD) reactor using the VLS-growth method and nanoimprint-patterned gold islands as a catalyst.[12,16,17] The untapered growth of the 100 nm Ge core NWs was performed at 320 °C using Monogermane ($GeH_4$) precursor, followed by the GeSn shell growth at 300 °C with the additional supply of Tin-Tetrachloride ($SnCl_4$) and Hydrogen Chloride (HCl) precursors. The effect of the $SnCl_4$ precursor flow on the morphology of the GeSn shell is shown in Fig. 1. A fixed growth time of 2 h was used in combination with a Ge/Sn ratio in the gas phase ranging from 1285 to 300. We note that when the $SnCl_4$ precursor is introduced in the CVD reactor the axial NW growth is suppressed, while the radial GeSn shell growth is promoted.[12,17] An increase in the diameter of the core/shell NWs is visible with increasing (decreasing) supply of the $SnCl_4$ precursor (Ge/Sn ratio), as visible in Fig. 1b-d. For Ge/Sn=1285 a more complex faceting of the GeSn shell morphology is observed (Fig. 1b), which becomes more prominent at a lower Ge/Sn ratio of 450 (Fig. 1c). Interestingly, when the $SnCl_4$ precursor flow is increased further (Ge/Sn=300) large Sn droplets are observed on the NW sidewall (Fig. 1d), as discussed in more detail later in the text. The $SnCl_4$ flow is not only a crucial parameter to control the thickness of the GeSn shell, but it also has a strong effect on the segregation of Sn and, in turn, on the NW morphology. A detailed insight into the evolution of the thickness and morphology of the GeSn shell is obtained using energy-dispersive x-ray spectroscopy (EDX) compositional maps acquired in cross-sectional scanning-TEM (STEM) (Fig. 2). In the Ge/Sn=1285 sample (Fig. 2a) a 20-30 nm thick GeSn shell terminated by six wide {112} facets connected by small rounded corners, possibly corresponding to six {110} nano-facets, is visible around the 100 nm Ge core. The EDX line-scans acquired along the radial <112> and <110> directions are shown in Fig. 2b, with a Sn content of ~8 at.% that is estimated on the main {112} facets. In addition, Sn-poor triangular regions (~4 at.%) are observed along <110>, which are likely related to initial {110} facets inherited from the shape of the Ge core (see below), that disappear during the shell growth. When the $SnCl_4$ flow is increased (Ge/Sn=450) a thicker 65±5 nm shell with Sn content up to 12-13 at.% along the radial <112> directions and 7-10 at.% along the radial <110> directions is obtained (Fig. 2c-d). Sn-poor triangular regions along <110> extend into the inner part of the shell, narrowing down to a sunburst-like geometry, similar to what is observed when using a 50 nm Ge core.[12,16] However, a more inhomogeneous morphology of the cross-section of the NWs is present when growing on 100 nm Ge cores, which is induced by the increased strain in the NWs[17] and by the higher $SnCl_4$ supply in the gas phase.

Indeed, an even larger change in the morphology of the GeSn shell is observed at the highest $SnCl_4$ flow (Ge/Sn=300), where enhanced segregation of Sn leads to phase separation with the formation of Sn droplets at multiple positions along the sidewall (Fig. 1d). The cross-sectional EDX map in Fig. 2e was acquired on a segment of the NW without Sn droplets. A 12-fold shape of the cross-section is observed with alternating Sn-rich and Sn-poor regions. As indicated in Fig. 2f, the Sn content increases from 8 at.% up to 18 at.% along the <112> direction, while a lower Sn content of 5-9 at.% is observed along the <110> direction.

The plot of the radial growth rate estimated along the <110> and <112> directions of the GeSn shell as a function of the Ge/Sn precursor supply is shown in Fig. 3. Two different data sets for 100 nm (from Fig. 2) and 50 nm Ge cores (Refs.[12,16]) are plotted. At a fixed GeH$_4$ flow, the GeSn radial growth rate increases with the SnCl$_4$ flow, and it becomes 4x times faster when moving from the low-Sn case at Ge/Sn=1285 to the richest one at Ge/Sn=300 grown on the 100 nm Ge cores, which indicates that vapor-solid growth occurs in a Sn-limited regime. The same trend is observed when using 50 nm cores. In addition, for a Ge/Sn ratio in the 450-550 range a higher growth rate is obtained around 50 nm cores compared to 100 nm cores, which most likely results from the reduced amount of strain in the shell during growth around the thinner Ge NWs. We highlight that the increased GeSn growth rate with the SnCl$_4$ precursor flow is similar to what is observed in conventional planar GeSn growths, where the growth kinetics are controlled by the SnCl$_4$ supply and the growth temperature, thus suggesting a co-decomposition mechanism with the GeH$_4$ precursor.[20,21] Interestingly, in the case of Ge/Sn=300, the growth rate along the <112> direction is ~20 % faster than along the <110> direction, resulting in the dodecagonal shape. The transition from a 6-fold to a 12-fold symmetry of the NW cross-section will be explained below in terms of the increased shell growth rate.

The increased amount of Sn supplied during the GeSn shell growth not only affects the growth rates, but it also increases the number of Sn droplets on the NW sidewall, as highlighted in the STEM image in Fig. 4a of a NW grown using a Ge/Sn=300. The associated EDX maps in Fig. 4b-c show the presence of Sn-rich droplets located not only on the shell surface, but also inside in the GeSn shell. Cross-sectional EDX (Fig. 4d) measurements performed on a portion of the NW with a droplet shows a Sn content of ~99.7 at.% Sn in the droplet, with the remaining ~0.3 at.% Ge signal being at the resolution limit of the EDX analysis. In addition, the same orientation-dependent Sn incorporation is observed in the shell as in Fig. 2e, but a depletion of Sn in the outer 20-40 nm of the GeSn shell is visible in Fig. 4d (dashed lines region). Thus, when the Sn droplet forms on the sidewall, the shell growth continues and Sn atoms diffuse to the droplet rather than being incorporated in the growing GeSn shell. Furthermore, the presence of the Sn droplets does not compromise the metastable state of the GeSn shell,[12] since the bulk diffusion of Sn atoms from the inner portion of the shell toward the surface is negligible.[8,9,11,22,23] As a result, the Sn segregation on the NW sidewall seems to have a less severe effect on the structural quality of the NW samples when compared to planar GeSn growth, where the presence of liquid Sn-rich droplets (above the Ge-Sn eutectic temperature of 231 °C) induces phase separation of the strained GeSn layer underneath.[24] The cross-sectional TEM and corresponding Fast-Fourier transform (FFT) images for the shell and droplet regions are shown in Fig. 4e, which indicate the presence of the β-Sn phase.[25]

**Theoretical growth model.** We first focus on the modeling of the shape transition from the dodecagonal cross-section observed at low Ge/Sn ratio (Fig. 2e) to the hexagonal one at high Ge/Sn ratio (Fig. 2a,c). Since the Ge cores expose six {112} and six {110} facets with similar lateral extension,[17] any change in the shell morphology occurring during the growth of the GeSn shell stems from the competition between the <112> and the <110> growth fronts. In the following description, limited to the two-dimensional (111)-cross section of the NW, such growth fronts will be modeled as {112} and {110} facets even if they can in principle result from a combination of different facets, especially in the case of low Ge/Sn ratio. In a minimal model, we consider the GeSn growth process as resulting from the combined effect of deposition from the precursors in the gaseous phase and redistribution of adatoms by surface diffusion. An isotropic distribution of the incoming material is assumed. The movements of adatoms formed at the surface follow the local chemical potential $\mu$. Diffusion is effective over a distance of the order of the diffusion length $\lambda$, and it is determined by the square root of the ratio between the mobility $M$ and the growth rate $\Phi$, *i.e.*: $\lambda \sim \sqrt{M/\Phi}$. As the experiments show a significant variation in the growth rate as a function of the Ge/Sn ratio (Fig. 3), material redistribution is expected to play a key role in the formation of the six-fold symmetric shell. The shape transition can then be rationalized by considering diffusion from {112} to {110} facets, as a consequence of a lower $\mu$ on the {110} facets than on the {112} ones. Such a significant difference is not justifiable by energetic arguments. Surface energies of {112} and {110} for pure Ge are indeed almost identical[26] (Supporting material S1) and, they are not expected to change significantly, particularly during the first stages of growth where the Sn content is low. Since the Sn incorporation in the shell is well above the ~1 at.% equilibrium solubility limit in Ge, the growth kinetics play a crucial role in the out-of-equilibrium GeSn deposition. However, an accurate characterization of the dynamics of Sn incorporation at the microscopic level would be extremely complex, requiring a detailed analysis of many atomistic processes, including the variety of chemical reactions involving the gaseous precursors. Nonetheless, the combined effect of these processes can be condensed into an effective parameter, the adatom incorporation lifetime $\tau$, accounting for the net rate of adatom incorporation on the different facets. As discussed in Refs.[27,28], a growth model including the incorporation kinetics can be setup by considering $\mu = \mu_{eq} + \tau v$. $\mu_{eq} \sim \kappa \gamma$ is the thermodynamic contribution to the adatom chemical potential, here assumed proportional to the local curvature $\kappa$ with isotropic surface energy density $\gamma$, and strain is neglected. $v$ is the profile velocity itself and is determined by deposition and surface diffusion: $v = \Phi + \nabla_S \cdot [M \nabla_S \mu]$. $\tau$ can be defined as a continuum function of the surface orientation,[28,29] with local maxima in correspondence of the observed facets. Indeed, in a kinetic growth regime, the (convex) crystal shape consists of those facets growing at a slower rate (*i.e.* high $\tau$).[30] For the present 2D case study, twelve local maxima (one every 30°) are imposed, corresponding to the set of <112> and <110> directions in the (111)-cross sectional plane of the NW. Following the previous considerations on the shape transition, $\tau$ maxima in the <112> directions are set greater than those along <110> ones. A factor of $\sim 2$ is estimated (a posteriori) to return a satisfactory match between simulation results and experiments (see below). A computationally efficient phase-field model is exploited to implement the evolution equations (see Methods section for details).

Figure 5 reports sequences of profile evolution obtained by simulations for three different ratios of mobility over growth rate ($M/\Phi = 1, 10, 50$) starting from a 100 nm dodecagonal Ge core. If deposition is fast enough to frustrate inter-facet diffusion (Fig. 5a) a dodecagonal facetted shape, formed by both {112} and {110} facets, is obtained as adatoms distribute over a short-range, which is only sufficient to keep the facets straight as corresponding to local maxima in $\tau$. Then, by increasing the $M/\Phi$ ratio (*i.e.* the diffusion length $\lambda$) the material exchange between the {112} and {110} facets is enabled and the slowest growing {112} facets (long- $\tau$) prevail over the {110}, converting the initial dodecagonal shape imposed by the core into one hexagon (Fig. 5b-c). Since the shape transition occurs by diffusive dynamics, it requires a finite time to complete depending on the relative adatom mobility $M/\Phi$, as evident by the different evolution paths of the central and right cases of Figure 5.

So far, we showed how the shape transition in the GeSn shell can be explained by considering different adatom incorporation kinetics on the two facets, without distinguishing between Ge and Sn atoms. However, experiments show that the morphological evolution occurs in combination to a non-uniform Sn distribution, consisting into a net enhancement of Sn content at the <112> growth fronts with respect to the <110> ones (Fig. 2). In order to understand the origin of this difference in Sn incorporation, we compute the energy variation resulting by the exchange of Ge atoms at the shell surface with Sn atoms from the gas. The change in the total energy of the crystal is estimated by *ab-initio* thermodynamics, based on the Density functional theory (DFT). Crystal slabs are cut along both {112} and {110} planes for both pure-Ge and GeSn at 9 at.% Sn. The former, illustrated in Fig. 6a, is considered as representative of the pristine growth stages while the latter corresponds to a later stage where a few GeSn layers have already been deposited on the Ge core. One of the Ge atoms is then replaced by Sn and the difference in energy with respect to the previous state $E - E_0$ is calculated. The procedure is repeated for every nonequivalent Ge site in the cell in order to select the energetically most favorable one. For the sake of simplicity, the selection was restricted only to the topmost layer of both {112} and {110} slabs. The optimal configuration for both facets is reported in Fig. 6b, while a site-dependent analysis can be found in the Supporting Information S2. As the exchange process involves Ge and Sn atoms in the gas phase, with energy $\mu_{Ge}^g$ and $\mu_{Sn}^g$ respectively, in order to estimate the actual energy gain due to the Sn incorporation, the total variation of energy $E_i = (E + \mu_{Ge}^g) - (E_0 + \mu_{Sn}^g)$ has to be considered, as sketched in Fig. 6a. This causes the result to depend on the partial pressures of both components, namely of their precursors, which are indeed tuned in the experiment. The plot in Figure 6c shows the variation in $E_i$ as a function of $\mu_{Sn}^g$ with respect to its bulk $\mu_{Sn}^b$, with the assumption of $\mu_{Ge}^g = \mu_{Ge}^b$ (see Method section for details). As evident, the {112} facet is predicted to have the lowest incorporation energy for any growth condition for both pure-Ge and GeSn slabs. For obvious reasons, this preference, tested for the initial layers, holds true for the subsequent shell growth, even if a more complex faceting may occur, so that we conclude that Sn is introduced more easily into the {112} fronts than the {110} fronts. Microscopically, this could be explained by noting that Ge-Ge bonds on the {112} surface are very stretched with respect to the bulk, in contrast with the bonds on the {110} surface. Thus, the replacement of a Ge atom by a Sn one is more convenient on the {112} facets, since the Ge-Sn bond length is larger than a Ge-Ge bond (see Supporting Information S2). As a result, {112} facets will grow with a higher Sn content than {110} facets, in agreement with the EDX results on all samples (Fig. 2).[17] The negative slope of $E_i(\mu_{Sn}^g)$ also indicates that the incorporation of Sn becomes more favorable when considering Sn rich growth conditions (*i.e.* low Ge/Sn ratio). In the opposite case, when $\mu_{Sn}$ is decreased below the bulk value, the energy $E_i$ becomes positive and Sn incorporation is not favored anymore. It is worth noticing, that the present analysis just states the energy advantage of placing a Sn atom at the surface with respect to leaving it in the gaseous phase, without considering either the reaction mechanism by which the process occurs, nor the kinetic path for it. Moreover, the option of forming liquid Sn droplets as in the experiments at high Sn fluxes is not explicitly considered and would introduce an upper limit to µ$_{Sn}$ with respect to the plotted range.

Let us finally combine the kinetic description of the shape transition and the atomistic explanation of the different Sn content along the <112> and <110> directions into a comprehensive model, tackling both the shape and composition evolution during the shell growth. To keep the description as simple as possible and limit the number of parameters, we start from the kinetic growth model employed for Fig. 5 and extend it to explicitly cope with the two alloy components, *i.e.* Ge and Sn. The overall model concept is reported in Ref.[29] and follows the seminal idea proposed by Tersoff in Ref.[31] of limiting intermixing effects to only within a few atomic layers from the free-surface.[32] Deposition and diffusion take place at the surface for each component $i$, *i.e.* $v_i = \Phi_i + \nabla_S \cdot [c_i M \nabla_S \mu_i]$ (same mobility $M$ is assumed) and their combination determines the advancing of the growth front ($v = \sum v_i$) as well as the change in the surface composition ($\partial c_i / \partial t = v_i - c_i v$). In the region underneath the surface the composition profile remains frozen-in, as bulk diffusion is suppressed. When neglecting mixing enthalpy, the chemical potential for each component will only include an additional term from configurational entropy, *i.e.* $\mu_i = (\mu_{eq} + \tau v) + kTlnc_i$, with $k$ the Boltzmann constant and $T$ the temperature. The model parameters are then set identical to those of Fig. 5, but for the flux of Sn $\Phi_{Sn}$ onto the surface. This is modulated as a function of the local profile orientation to include, in an effective way, the preference of Sn to stay on {112} surfaces, as indicated by experiments and atomistic simulations. A simple sinusoidal variation is considered with maxima along the <112> directions and minima in the <110> ones, resulting in a realistic Sn content of ~10 at.% on {110} facets and ~ 15 at.%, respectively.

Including the compositional field into the aforementioned phase-field approach enables us to simultaneously trace the advancing of the shell growth front and the Sn segregation effects. As already discussed for the shape transition (Fig. 5), also in this coupled dynamics, the key is the relative role of diffusion and deposition, *i.e.* the mobility/growth rate ratio $M/\Phi$. In Figure 7 (see also Fig. S4), we show three simulated profiles obtained by varying the $M/\Phi$ ratio in order to match the EDX maps of Fig. 2. The Sn content is mapped by colors and its radial variation along both <112> and <110> directions is plotted. The correspondence between simulation and experiments is compelling as most of the key features are reproduced. In particular, in the case of Fig. 7a, where adatom redistribution is quite limited, we achieve a dodecagonal shape with <112>-Sn-rich sectors and <110>-Sn-poor ones. A slight concavity in the <110> direction is also distinguishable, as trace of the frustrated diffusion from the {112} regions toward the {110} regions. In the case of long-range diffusion (center and right Fig. 7b-c), the kinetically-preferred hexagonal shape is obtained right after deposition of a thin dodecagonal shell. The progressive shrinkage of the {110} facets, originating from the Ge core, is well marked by a Sn-poor triangular region which extends up to a certain shell thickness determined by the $M/\Phi$ ratio. As indicated in the plot of the composition along the <110> radius, the Sn content within the {110} sector is not constant but increases with the shell thickness. This is explained by the arrival of material diffusing from the Sn rich {112} regions into the {110} region (see Fig. S4), mixing with the Sn-poor adatoms deposited on the smaller {110} area.

It is worth noting that in the present work the systematic analysis of shape and composition was focused on samples with 100nm cores, quite large compared to the ones typically discussed in literature.[12,13,16,33] This was essential in order to have dodecagonal shapes. Indeed, in all other samples with smaller core radius (down to 50 nm) only a {112}-bounded hexagonal cross-section was observed for all Ge/Sn ratio (unless irregular, defected shapes where obtained).[17] This is compatible with the model as the shape transition depends on the relative extent of diffusion length with respect to the facet dimension. By reducing the core size it is then reasonable to expect that the kinetic shape is accessible by lower $M/\Phi$ ratio than required for the larger cores considered here. An example of the influence of the core size on the final shape of the resulting NW is shown in Figure 8. Even if theoretically, a dodecagonal structure on the 50 nm core could still be obtained by lowering further $M/\Phi$, the relevant growth regime may not have been explored in growth experiments yet.

In the present simulations the appearance of nm-thin, Sn-poor (1 at.% lower than facets) stripes is observed along the <110> directions. This resembles the sunburst-like structure commonly observed in nanowires. In order to achieve the localized Sn segregation observed in experiments (~5% lower than facets) additional effects, *e.g.* mobility differences between Ge and Sn adatoms, need to be included in the model.

**Optical properties.** The effect of the incorporation and segregation of Sn on the optical properties of the NWs is shown using low-temperature (4 K) photoluminescence (PL) measurements.[12] No PL emission was observed in the Ge/Sn=1285 sample, which is most likely due to the low Sn incorporation (~8 at.%) in combination with the highly compressively-strained nature and reduced volume (20-30 nm-thick) of the $Ge_{0.92}Sn_{0.08}$ shell. To promote an indirect to direct band gap transition in unstrained GeSn the incorporation of Sn should be higher than 8-9 at.%, which is not achieved in the Ge/Sn=1285 sample. No optical emission was also detected in the Ge/Sn=300 sample (Fig. 4), thus demonstrating that the severe Sn segregation and phase separation compromise the quality of the heterostructured NWs. The PL spectra for the samples grown with Ge/Sn=450-600 are shown in Fig. 9a. In the Ge/Sn=600 sample (dark red curve) a single PL peak is observed at 0.44 eV (*i.e.* 2.8 μm wavelength) with a FWHM of ~80 meV. By increasing the $SnCl_4$ flow to Ge/Sn=450 (blue curve) the optical emission shifts to 0.41 eV (*i.e.* 3.0 μm wavelength) with a FWHM of 60-70 meV. The red-shift of the PL peak with increasing $SnCl_4$ supply (*i.e.* enhanced Sn incorporation) is in agreement with what was observed for Ge/GeSn core/shell NWs grown using a 50 nm core and Ge/Sn=518-740 from Ref.[12], which are also plotted as a comparison in Fig. 9a. The higher, non-uniform Sn content (18-20 at.%) in the shell grown around the 100 nm cores compared to the (~13 at.%) 50 nm NWs reduces the energy of the emitted PL, in agreement with the recent studies on GeSn thin films and nanostructures.[2,5,14] However, the emission energy of GeSn is also influenced by the higher residual compressive strain in the shell when grown around the thicker 100 nm Ge cores,[17] which increases the band gap value, thus in the opposite direction compared to the shift induced by a higher Sn incorporation. In addition, lower intensity and a broader emission peak is obtained for the NWs grown using the 100 nm Ge cores instead of 50 nm cores,[12] which would suggest reduced optical recombination in presence of residual compressive strain and non-uniform Sn incorporation in the GeSn shell. We note that the PL emission at 4 K is most likely dominated by band gap-related localized states, such as alloy fluctuations, impurities or defects. To confirm this, PL measurements performed in the 4-100 K temperature range on the 100 nm Ge core and Ge/Sn=450 shell sample (Fig. 9b-c) show a quenching of the optical emission at ~100 K. As a comparison, in the thinner 50 nm Ge core samples the PL emission is observed until 300 K, with a thermal detrapping from the localized states into the direct band gap emission above 100 K.[12] In Ge/GeSn core/shell NWs grown using a 100 nm Ge core the higher amount of compressive strain in the GeSn shell combined with the enhanced Sn segregation on the {112} facets of the shell lead to a rapid quenching of the PL emission with increasing temperature, thus indicating the limited optical properties of these core/shell NWs when compared to the ones grown using the 50 nm Ge core. Similarly, room-temperature PL emission was also promoted in strain-free GeSn NWs[14] and in GeSn thin films when under-etched in free-standing microdisks.[34] Therefore, major care is required in minimizing strain and segregation in GeSn to enhance the optical properties at mid-infrared wavelengths while preserving room-temperature operation.

## CONCLUSION

The growth of metastable GeSn semiconductors in a Ge/GeSn core/shell nanowire geometry has been investigated in-depth by combining experimental and theoretical analyses. We show that by increasing the supply of Sn atoms in the gas phase during growth, the morphology and composition of the GeSn shell is strongly affected. A 6-fold to 12-fold change in the symmetry of the NW cross-section by increasing the size of the six <110> corners in the GeSn shell with respect to the (main) six {112} facets is observed. Enhanced segregation of Sn is observed at higher Sn supply with an increasing compositional difference between the Sn–poor <110>-oriented stripes and Sn-rich <112>-oriented ones, eventually leading to phase separation and nucleation of Sn droplets on the NW sidewall.

In order to understand the GeSn growth mechanism at the nanoscale a multi-scale approach was exploited and it reveals a deep connection between the Sn segregation effects and the morphological changes in the NW shell for different growth conditions. We predict that {110} facets (if present) incorporate a lower amount of Sn compared to {112} facets at any supply of Sn atoms in the gas phase during growth. Indeed, under the assumption that adatom incorporation on {112} facets is slower than on {110} facets our model reproduces the experimentally-observed shapes and composition profiles with high accuracy. The interplay between facet dependent incorporation and surface adatom kinetics is inferred to be the key process that determines both segregation of Sn and the morphology of the GeSn shell. Dedicated studies on the microscopic processes, especially on the complex surface chemistry processes of the GeSn precursors, will help to assess the origin of the kinetic effects. Interestingly, despite the simplifications made in our model, the developed 2D model provides an accurate description of the experimental findings, although it does not cover the whole complexity of the GeSn growth dynamics like misfit strain and plastic relaxation which might occur for large cores.[17] We conclude that these effects do not significantly affect the growth dynamics. In addition, we highlight the effect of strain and Sn segregation in reducing the optical emission of the GeSn shell when using a 100 nm Ge core compared to 50 nm. The 4 K photoluminescence emission in the thicker cores is dominated by localized states, with a quenching of the emission above 100 K, while in the thinner core band-to-band recombination is observed at room temperature. These findings clearly show that minimizing strain and segregation in the Ge/GeSn core/shell system is of paramount importance to preserve the high optical quality of these NWs at mid-infrared wavelengths.

## METHODS

**NW growth and characterization.** The VLS-growth of Au-catalyzed arrays of Ge/GeSn core/shell NWs was performed in a Aixtron CCS chemical vapor deposition (CVD) reactor using $H_2$ as a carrier gas, following the growth protocol recently established in Ref.[12]. Monogermane ($GeH_4$), tin tetrachloride ($SnCl_4$), and hydrogen chloride (HCl) were used as precursors for the NW growth using $H_2$ as a carrier gas. The array of Au droplets with a diameter of 100 nm and a pitch of 500 nm was fabricated on a Ge (111) substrate using nanoimprint lithography. First, the nucleation of the Ge NWs was performed at 425°C with the supply of the $GeH_4$ precursor and at a reactor pressure of 100 mbar. Next, the sample was cooled down to 320°C and the untapered Ge NWs were grown at a reactor pressure of 75 mbar. Lastly, the GeSn shell was grown at 300°C for 2 hours and at a reactor pressure of 50 mbar. The $GeH_4$ and HCl precursors were supplied with a constant molar fraction of $4 \cdot 10^{-3}$ and $4 \cdot 10^{-4}$, respectively, while the $SnCl_4$ molar fraction was varied in the $3\text{-}10 \cdot 10^{-6}$ range, leading to a Ge/Sn ratio in the gas phase of 1285-300.
Transmission Electron Microscopy (TEM) studies were performed using a probe-corrected JEOL ARM 200F, operated at 200 kV, equipped with a 100 $mm^2$ Centurio SDD EDS detector.

**DFT.** First principle calculations are performed using density functional theory (DFT) and a planewave basis sets as implemented in the Quantum Expresso code.[35] The exchange-correlation is treated in the generalized gradient approximation (GGA) as parameterized by Perdew, Burke, and Ernzerhof (PBE) and projector augmented wave (PAW) pseudopotentials are used, setting a wavefunction cut-off of 60Ry. The {110} and {112} Ge surfaces are modeled by using symmetric slabs with 6 atomic layers (108 atoms), (3×2) and (3×1) in-plane periodicity, respectively, and considering their most stable reconstructions among different ones proposed in the literature.[36–39] The incorporation energy of Sn onto the Ge surface is calculated as: $E_i = (E + \mu_{Ge}^g) - (E_0 + \mu_{Sn}^g)$; where $E_0$ is the DFT total energy of a pure Ge slab while E is the energy of the corresponding lowest energy configuration with a Sn substituting a Ge atom on the surface. $\mu_{Ge}^g$ and $\mu_{Sn}^g$ are the chemical potential of Ge and Sn, respectively. The former is set to the chemical potential of bulk Ge, since it is expected that the conditions required to form crystalline Ge shell should bring the interface close to the equilibrium with the underlying Ge core. Contrary, the chemical potential of Sn has been varied with respect to its bulk reference, in order to mimic changes of the Ge/Sn ratio. Thus, in Figure 6(c) the incorporation energy is plotted as a function of the variation of the Sn chemical potential with respect to its bulk value.

**GeSn shell growth model.** The evolution of the GeSn NW shell during growth is investigated by a phase-field model coupling the kinetic description of the morphological evolution defined in Ref.[28] to the two component dynamics developed in Ref.[40] (see Supporting Information S3 for more details). The system is defined by two parameters: (i) the phase-field function φ, tracing implicitly the growth front as the diffuse-interface between the solid (where φ=1) and the surrounding vacuum phase (φ=0);[41] (ii) the composition field $c$ specifying the local Sn content. The profile evolution then results from the net flux of both components at a given point: $\partial \varphi / \partial t = \sum_i (\Phi_i |\nabla \varphi| + \nabla \cdot [M \nabla \mu_i])$, where $\Phi_i$ and $\mu_i$ are the growth rate and the surface chemical potential of the i-th (Sn or Ge) component. The mobility $M$ is restricted at the surface region by setting $M(\varphi) \sim \varphi^2(1-\varphi)^2$. At the same time, the unbalance in the flux of the two components is responsible for the eventual variation in the local composition $c$, as $\partial(c\varphi)/\partial t = \Phi_{Sn}|\nabla\varphi| + \nabla \cdot [M\nabla\mu_{Sn}]$. While Ge deposition is set with a uniform rate $\Phi_{Ge}$=0.875nm/s, the Sn growth rate varies as $\Phi_{Sn} = 0.125 + 0.025\cos(6\theta)$, with maxima corresponding to the <112> directions and minima in the <110> ones. The chemical potential of each component includes a (isotropic) surface energy and a kinetic term, undifferentiated for Sn and Ge for the sake of simplicity, and mixing entropy, i.e. $\mu_i = \gamma[-\epsilon\nabla^2\varphi + (1/\epsilon)B'(\varphi)] + \epsilon\tau(\partial\varphi/\partial t) + kT\ln c_i$, with a double-well potential $B(\varphi) = (18/\epsilon)\varphi^2(1-\varphi)^2$, $\epsilon$ defines the width of the $\varphi$ diffused interface for the phase-field description, $k$ the Boltzmann constant and $T$ the temperature, here set equal to 320°C in compliance with our experiments. The variation of incorporation times $\tau$ with respect to the profile orientation $\theta$ is obtained by the convenient formula of Ref.[29]. In particular, we set $\tau_{112}$=10 and $\tau_{110}$=5. The numerical solution of the evolution problem is performed by a finite elements method exploiting the AMDiS toolbox.[42]

**Photoluminescence measurements.** The optical properties of the NWs were probed using macro-photoluminescence (PL) measurements. The samples with as-grown wires were mounted in a vertically oriented helium flow cryostat with accurate temperature control. A 976 nm CW-laser set to 41 mW, focused down to a spot of ~35 μm in diameter using a using a 2.1 cm focal length off-axis parabolic mirror, was used for the excitation. The same mirror was then used to collimate the PL signal and couple it into a Thermo Scientific IS50r FTIR equipped with a liquid nitrogen cooled MCT detector. The thermal background radiation was minimized by modulating the laser at 35kHz and using a lock-in amplifier, while operating the FTIR in step-scan mode. The full optical path was purged with nitrogen to minimize absorption due to water and carbon-dioxide in air.

**FIGURES CAPTIONS**

**Figure 1.** (a-d) SEM images of the Ge/GeSn core/shell NW arrays (tilting angle 30 °) grown using a Ge core diameter 100 nm (a) and a GeSn shell with a Ge/Sn ratio of 1285 (b), 450 (c), and 300 (d).

**Figure 2.** (a-f) Cross-sectional EDX compositional maps and related line-profiles along the <112> and <110> radial directions of the NWs, acquired for Ge/Sn ratios of 1285 (a-b), 450 (c-d), and 300 (e-f). Enhanced Sn incorporation along the <112> radial direction of the GeSn shell is observed with increasing $SnCl_4$ supply.

**Figure 3.** Plot of the growth rate of the GeSn shell as a function of the Ge/Sn ratio, estimated from Fig. 2 (solid and hollow circles). The data points adapted from Refs.[12,16] are also shown (solid and hollow triangles).

**Figure 4.** (a) STEM image of a NW grown with Ge/Sn=300. (b-c) EDX compositional maps for Ge (b) and Sn (c) atoms showing phase separation with the presence of Sn droplets on the NW sidewall. (d) Cross-sectional EDX compositional map acquired on a spot where a Sn droplet is present. (e) Bright-field TEM image of the GeSn-Sn droplet interface acquired along the [-122] zone axis, tilted 15 ° from the main <111> axis to visualize lattice fringes in the Sn particle. Insets: corresponding FFT images of the Sn droplet and GeSn shell regions.

**Figure 5.** (a-c) Profile evolution from simulations with isotropic deposition and faster adatom incorporation along the <110> direction, as a function of the mobility/growth rate ratio $M/\Phi$ of 1

(a), 10 (b), and 50 (c). A regular dodecagon is set as initial profile, mimicking the Ge core shape and profiles are reported at three subsequent times. Colored sectors are shown to view the progressive evolution of the two facets along the <112> (red) and <110> (blue) directions.

**Figure 6.** (a) Representation of the simulation cell used for the *ab-initio* calculation of the incorporation energy for a {112} surface, comparing the state with or without a Sn surface atom. (b) Lateral and top views of the atomic structure of the {112} and {110} surfaces. A Sn atom (red) is positioned at the most favorable site. The unit cell is marked by a dashed rectangle. (c) Incorporation energy computed by *ab-initio* calculations of both {112} and {110} facets as a function of the Sn chemical potential.

**Figure 7.** (a-b) Profiles obtained by growth simulations performed for different mobility/growth rate conditions $M/\Phi$ of 5 (a), 30 (b), and 50 (c), mimicking the experimental cases in Fig. 2. Shell thicknesses are set to match the experimental ones. A magnification of one shell corner is reported for the case of slow deposition (c) to highlight the fast consumption of the {110} facet. The radial variation of Sn along both <112> and <110> directions is simulated up to a shell thickness of 100 nm.

**Figure 8.** (a-b) Simulation profiles obtained for identical mobility/growth rate conditions ($M/\Phi = 10$) and Ge core radius of 100 nm (a) and 50 nm (b).

**Fig. 9.** (a) PL spectra acquired at 4 K for Ge/GeSn core/shell NWs grown using 100 nm core and a Ge/Sn ratio of 450 (blue curve) and 600 (dark red curve) measured at an excitation power density of 1.9 kW/cm$^2$ (976 nm laser). The data on Ge/GeSn core/shell NWs grown using 50 nm core and a Ge/Sn ratio of 518 (red curve) and 740 (green curve) from Ref.[12] are also shown, measured at 7.8 W/cm$^2$ (405 nm laser). (b) PL spectra acquired in the 4-100 K temperature range for the sample

grown with Ge/Sn=450 and Ge core of 100 nm. (c) Integrated PL intensity as a function of the inverse of the temperature for the data in (b).

## ASSOCIATED CONTENT

**Supporting Information.**

The Supporting Information is available free of charge on the

ACS Publications website at DOI: ….

Additional information on the surface energy calculations, incorporation energy, and shell growth model.

## Preprint submission

Simone Assali, Roberto Bergamaschini, Emilio Scalise, Marcel A. Verheijen, Marco Albani, Alain Dijkstra, Ang Li, Sebastian Koelling, Erik P.A.M. Bakkers, Francesco Montalenti, Leo Miglio. Kinetic control of morphology and composition in Ge/GeSn core/shell nanowires. 2019, arXiv:1906.11694. Name of Repository. https://arxiv.org/abs/1906.11694 (accessed January 16th, 2020).

## AUTHOR INFORMATION

* corresponding authors: simone.assali@polymtl.ca , roberto.bergamaschini@unimib.it , emilio.scalise@unimib.it

¥: these authors contributed equally to this work.

## ACKNOWLEDGEMENTS

The authors thank P. J. van Veldhoven for the technical support with the MOVPE reactor and L. Gagliano for the nanoimprint lithography of the Ge substrate. This work was supported by the Dutch Organization for Scientific Research (NWO-VICI 700.10.441), and by the Dutch Technology Foundation (STW 12744) and Philips Research. Solliance and the Dutch province of Noord-Brabant are acknowledged for funding the TEM facility. We acknowledge the CINECA award under the ISCRA initiative, for the availability of high performance computing resources and support.

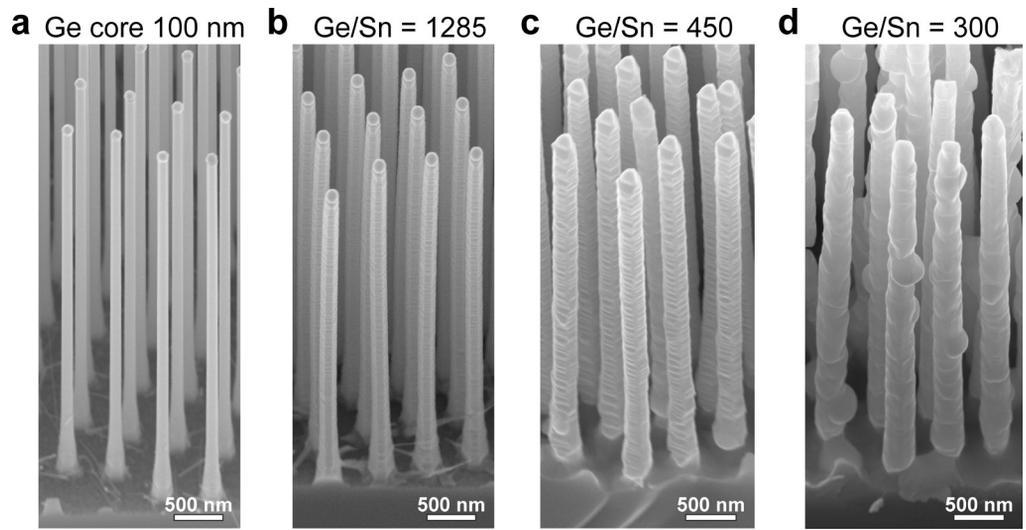

Figure 1

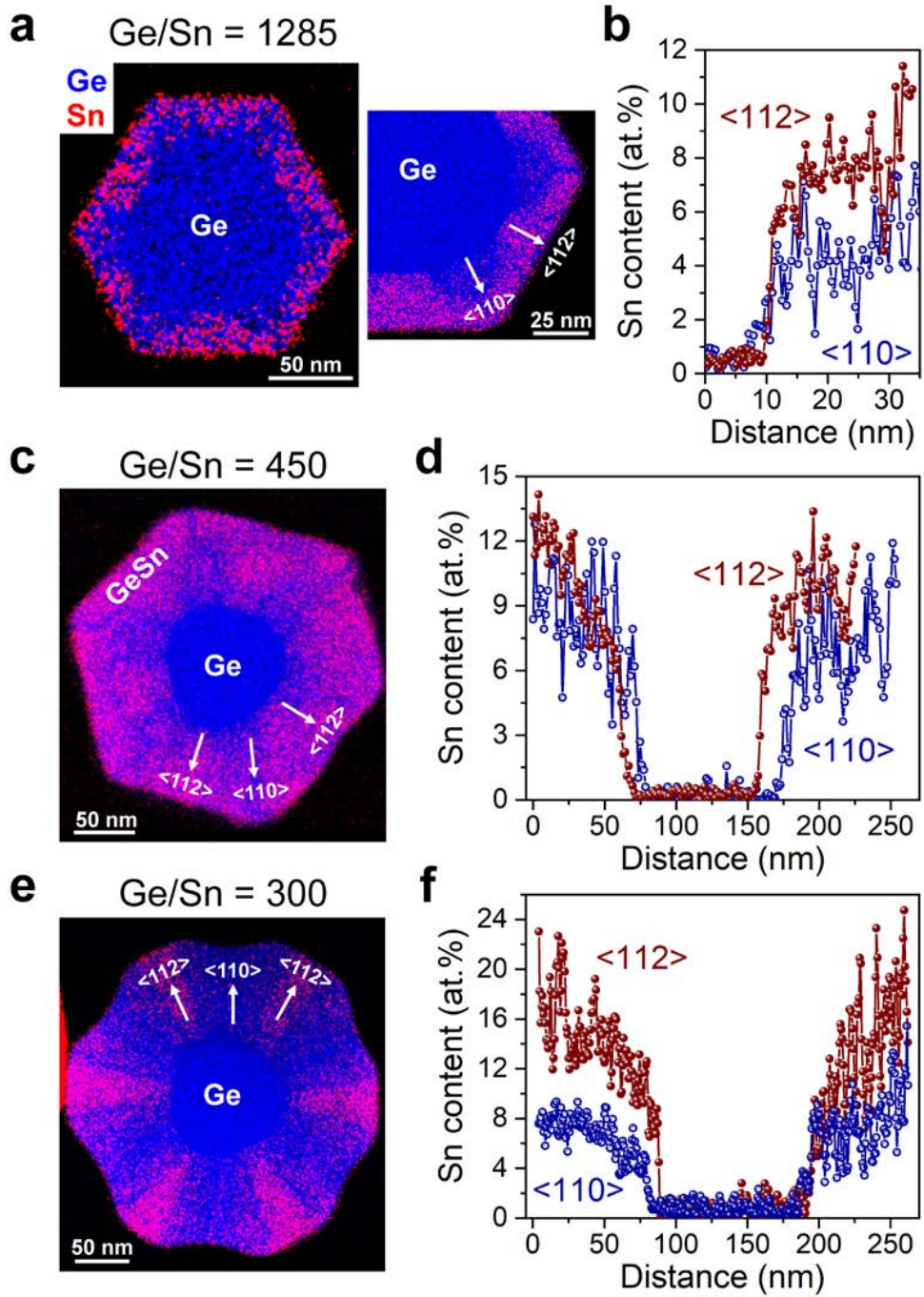

**Figure 2**

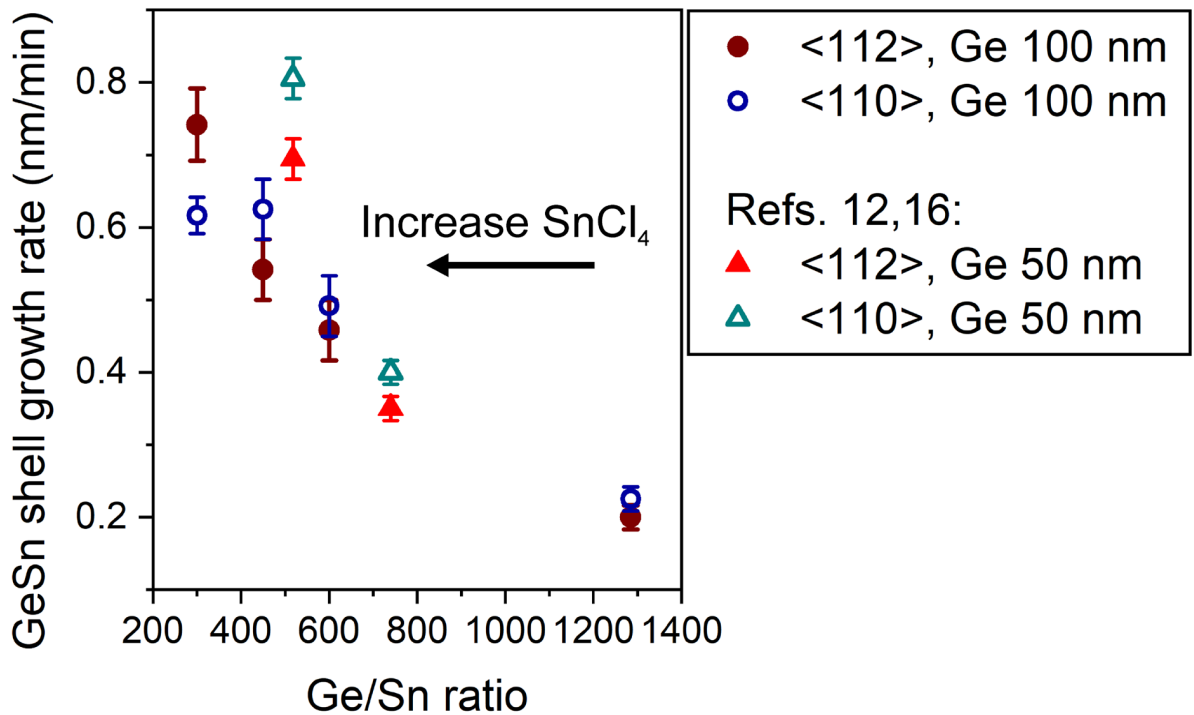

**Figure 3**

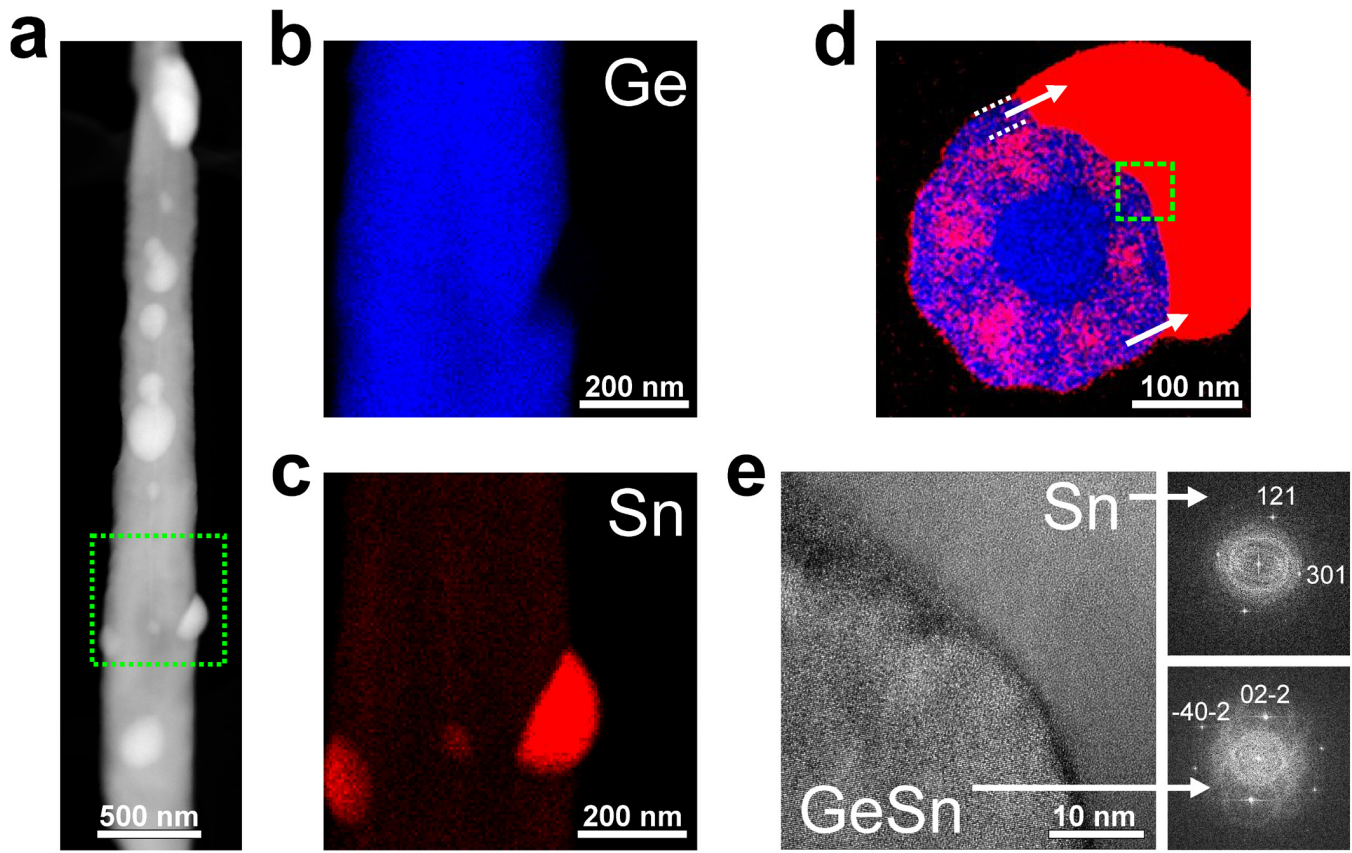

Figure 4

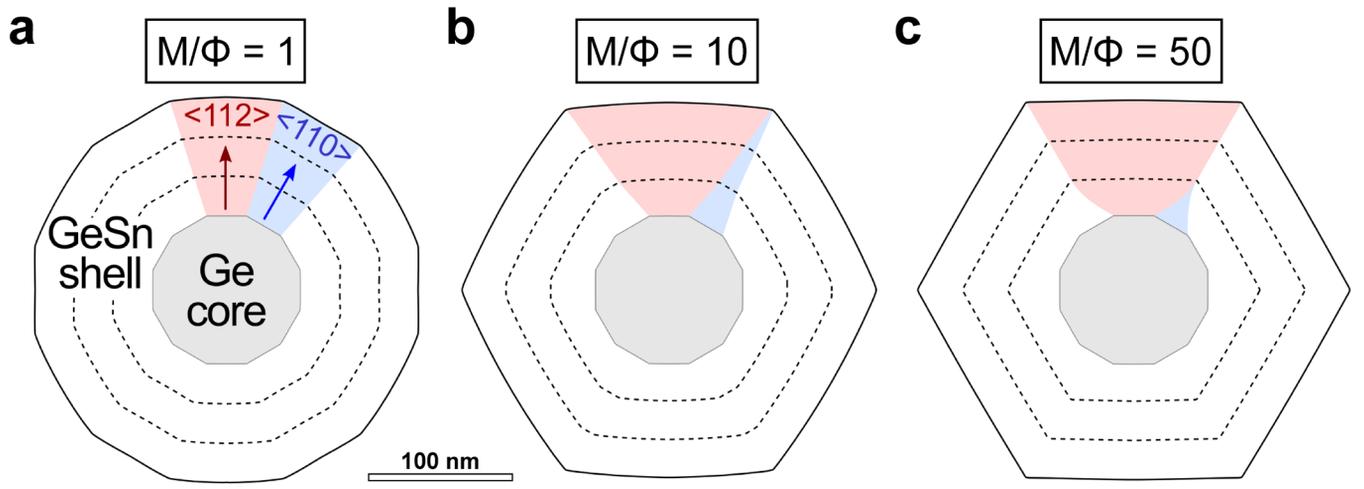

Figure 5

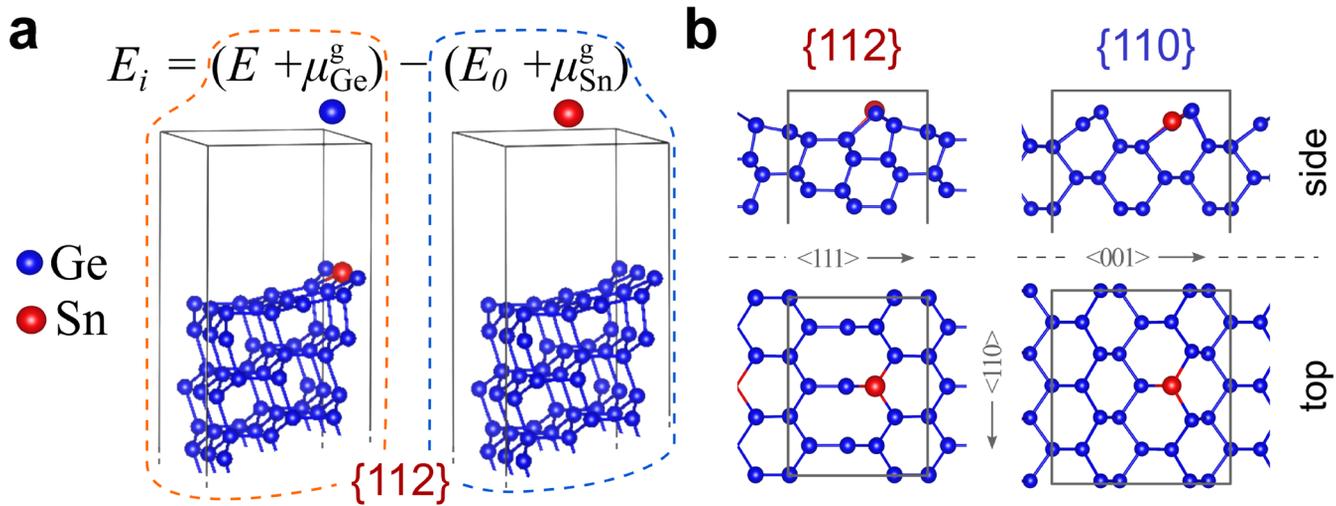
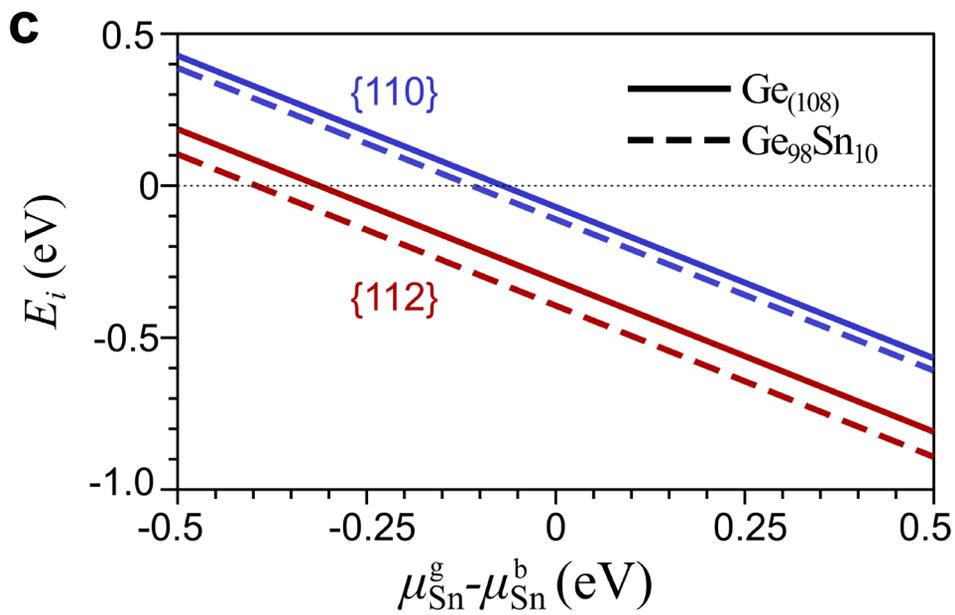

**Figure 6**

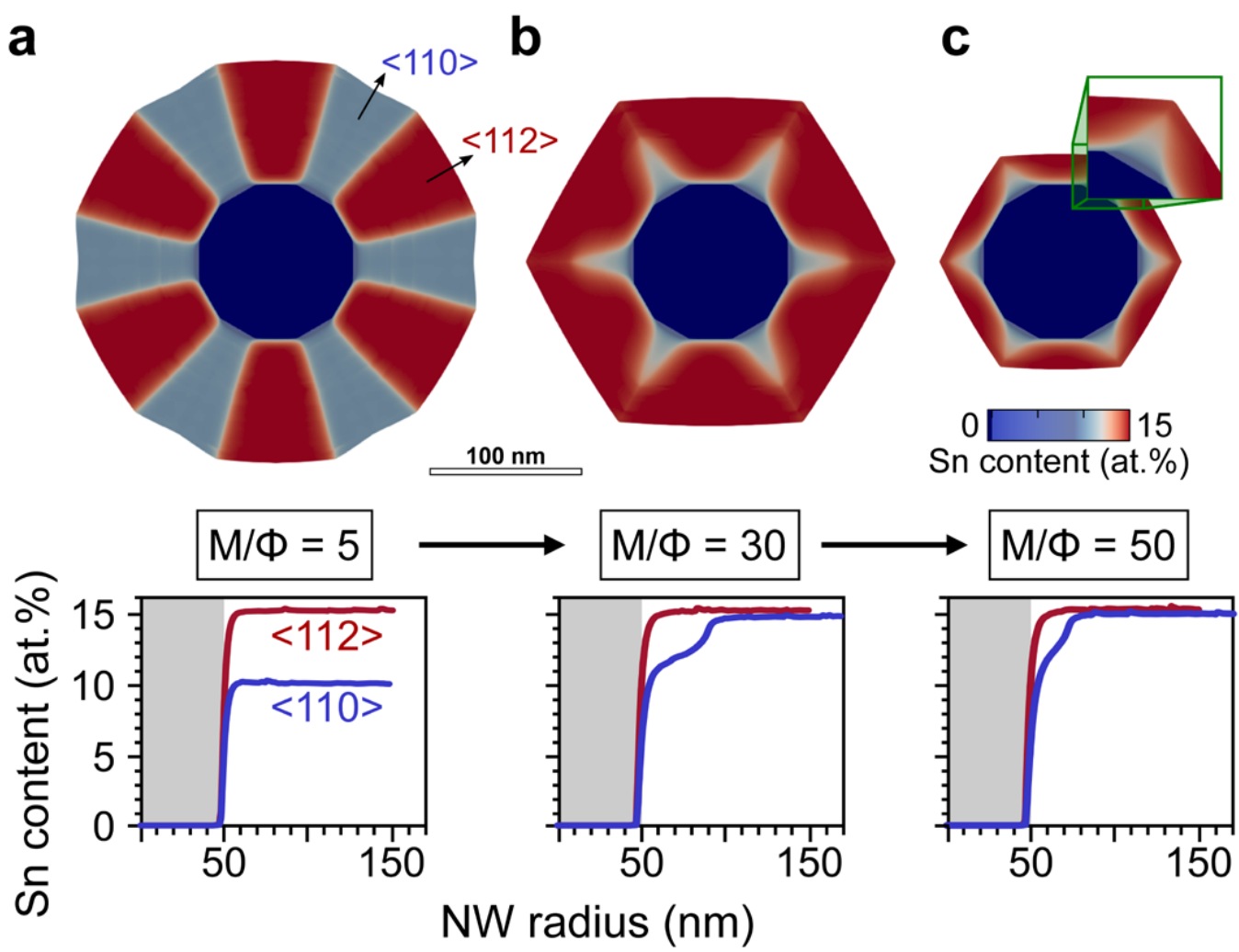

**Figure 7**

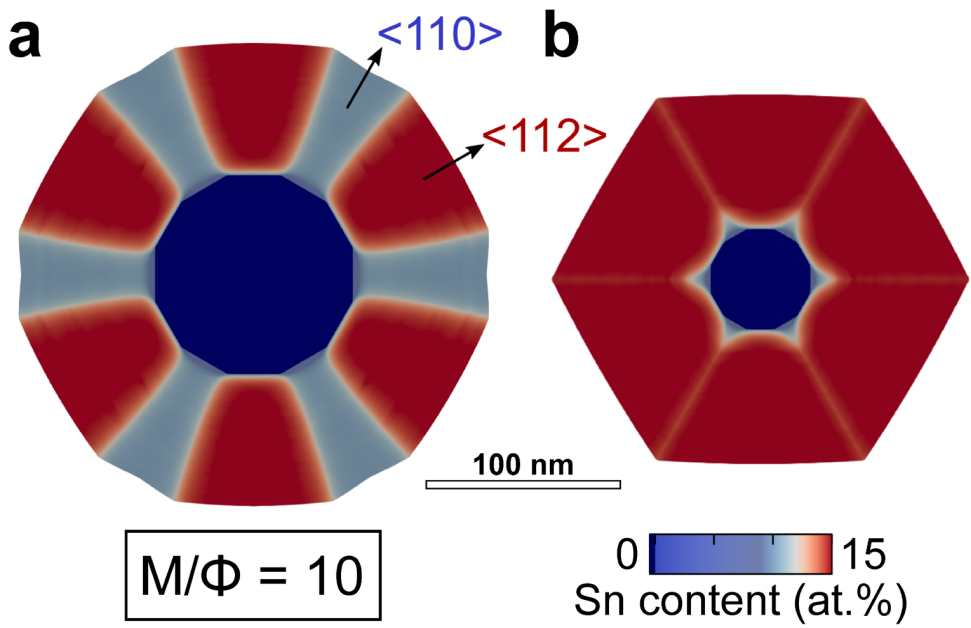

**Figure 8**

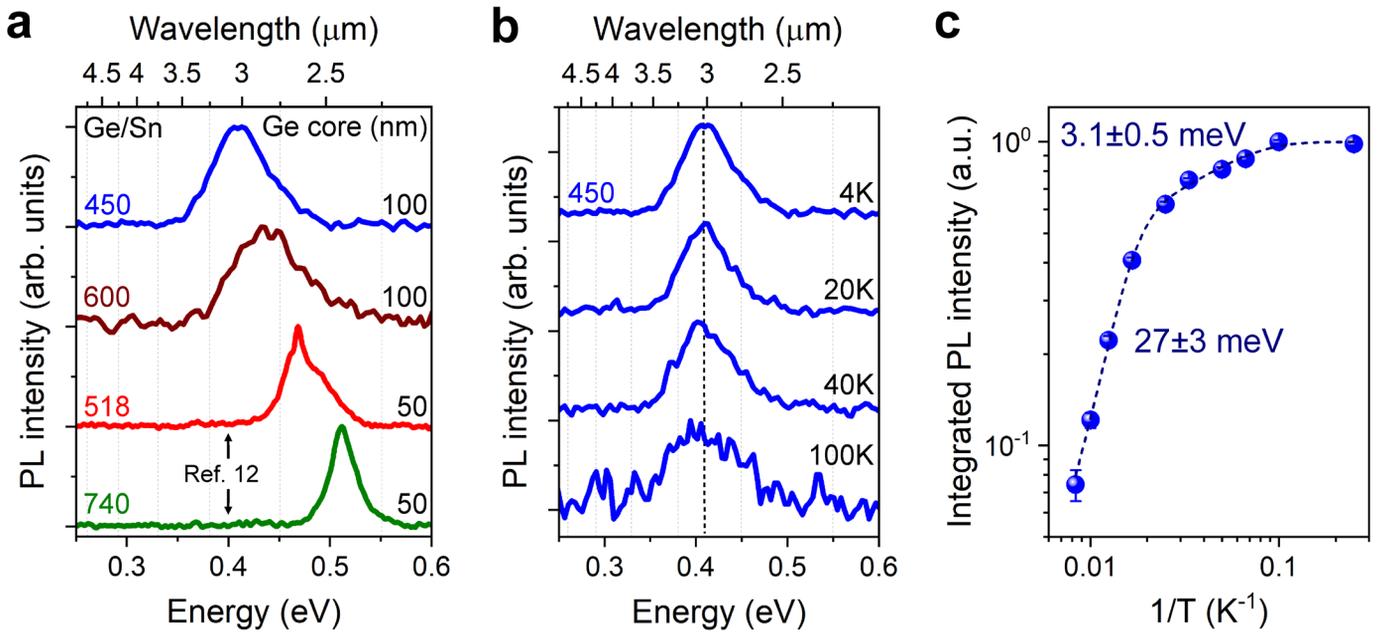

**Figure 9**

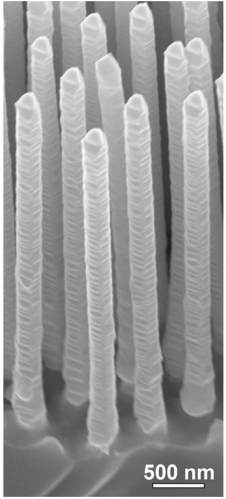
Ge/GeSn core/shell NWs

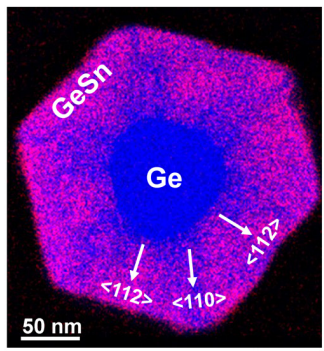
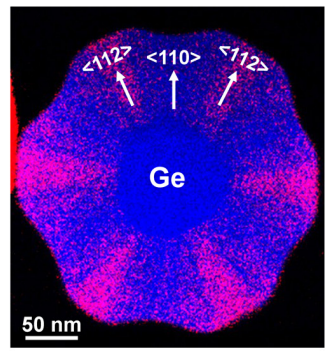

Increased SnCl$_4$ supply

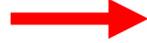
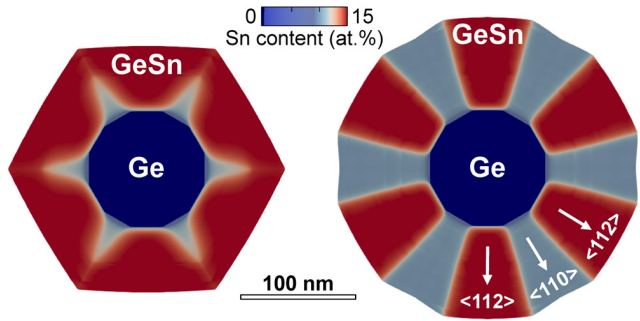

Table of contents

# Supporting information:

# Kinetic Control of Morphology and Composition in Ge/GeSn Core/Shell Nanowires

Simone Assali,[1,2,¥,*] Roberto Bergamaschini[3,¥,*], Emilio Scalise[3,¥,*], Marcel A. Verheijen,[4] Marco Albani[3], Alain Dijkstra,[1] Ang Li,[1,5] Sebastian Koelling,[1] Erik P.A.M. Bakkers,[1,6] Francesco Montalenti,[3] and Leo Miglio[3]

[1] Department of Applied Physics, Eindhoven University of Technology, 5600 MB Eindhoven, The Netherlands
[2] Department of Engineering Physics, École Polytechnique de Montréal, C. P. 6079, Succ. Centre-Ville, Montréal, Québec H3C 3A7, Canada
[3] L-NESS and Dept. of Materials Science, University of Milano Bicocca, 20125, Milano, Italy
[4] Eurofins Materials Science BV, High Tech Campus 11, 5656AE Eindhoven, The Netherlands
[5] Beijing University of Technology, Pingleyuan 100, 100124, P. R. China
[6] Kavli Institute of Nanoscience, Delft University of Technology, 2600 GA Delft, The Netherlands

# Contents

# S1. Surface energy calculations for Ge core {110} and {112} facets by DFT

| Facet | $\gamma$ (meV/Å$^2$) | |
|---|---|---|
| | Present work | From Ref. 1 |
| {110} | 59.8 | 60.0 |
| {112} - 1×1 | 59.5 | 60.0 |
| {112} - 2×1 | 62.3 | - |

Table S1. Surface energy of both {110} and {112} facets as obtained by DFT. Two different reconstructions are compared in the case of the {112} surface. Our results are compared with literature data from Ref. 1.

# S2. Incorporation energy

The energy required to incorporate a Sn atom from the gas phase into the surface can be quantified by comparing the total energy of two separate systems. The first one is made of a pure Ge lattice with a pure Ge surface, while the second one has a single Sn atom on the surface. This compares the variation in the free energy when a surface Ge atom is substituted with a Sn one from the gas, and the Ge free atom which is ideally removed from the surface is accounted by means of the chemical potential of the gas phase. In addition, we repeated the calculation for a GeSn slab, formed of 98 Ge atoms and 10 Sn atoms.

In details, the incorporation energy is defined as the difference between the total free energy for the two cases: $E_i = [E_{tot}^{Sn} + \mu_{Ge}] - [E_{tot}^{Ge} + \mu_{Sn}]$.

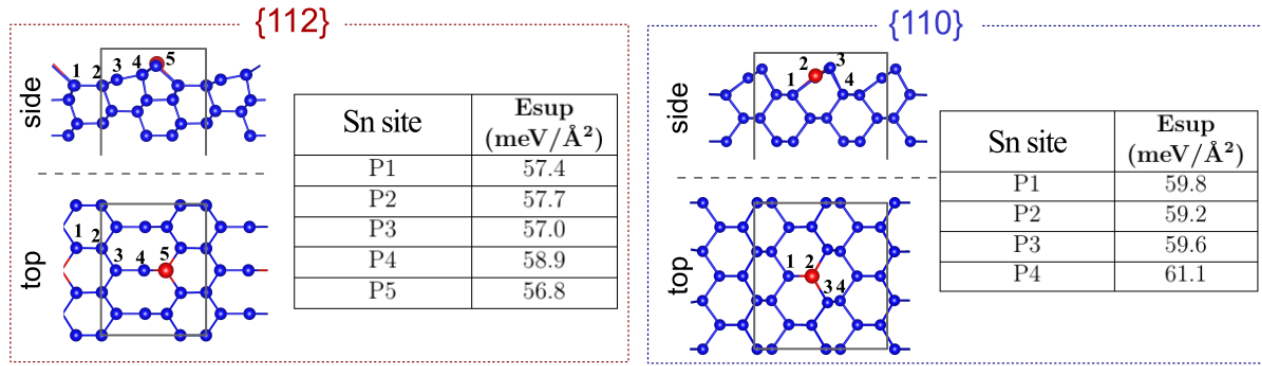

Figure S1: comparison of different Sn (red atom) incorporation sites for {112} and {110} facets.

The surface site for Sn incorporation has been selected by comparing the energy of the system for different positions of the Sn atom and by selecting the minimum energy configuration, as described in Fig. S1.

A microscopic interpretation for the different incorporation energy for the two facets can be based on the difference in the Ge-Sn bond length for the incorporated Sn atom, with respect to Sn-Sn bonds in pure Sn, as listed in Table S2. Interestingly, the difference between the Ge-Ge bond length in pure Ge and the Ge-Sn bond length is very small on the (112) surface but not on the (110) surface. This suggests that when Sn is introduced in the (112) surface it is less strained than on the (110) surface.

|  | Bond length (Å) |
|---|---|
| Ge-Sn (110) | 2.68 |
| Ge-Sn (112) | 2.82 |
| Sn-Sn (bulk Sn) | 2.88 |
| Ge-Ge (110) (pure Ge) | 2.57 |
| Ge-Ge (112) (pure Ge) | 2.79 |
| Ge-Ge (bulk Ge) | 2.49 |
| Ge-Sn (bulk $Ge_{98}Sn_{10}$) | 2.64-2.68 |

Table S2. Bond lengths on the (112) and (110) surfaces for Ge, GeSn, and Sn.

## S3. GeSn shell growth model – Technical details

The cross-section of the NW is traced implicitly by the phase-field function $\varphi = 0.5[1 - \tanh(3d/\epsilon)]$, with $d$ the signed-distance between a point $\vec{r}$ and the surface profile and $\epsilon$ the interface amplitude, here set equal to 4 nm. The volume of the NW corresponds to the region with $\varphi=1$ while the surrounding vacuum region is identified by $\varphi=0$. A diffuse interface of amplitude $\sim\epsilon$ returns a smooth transition between the two regions in correspondence of the NW free surface, nominally located at the $\varphi=0.5$ isoline. A composition field $c$ is used to define the Sn content within the NW. The system free energy $G$ is the sum of surface energy and entropy of mixing:

$$G[\varphi, c] = \int \gamma \left[\frac{\epsilon}{2}|\nabla\varphi|^2 + \frac{1}{\epsilon}W(\varphi)\right] dx +$$

$$+ \frac{kT}{V_a} \int [c \ln c + (1-c)\ln(1-c)]\, dx$$

where $W(\varphi) = 18\varphi^2(1-\varphi)^2$ is a double-well potential, $\gamma$ is the surface energy density (here assumed to be isotropic, a generalization to the anisotropic case can be found in Ref. 2), $k$ is the Boltzmann constant, $T$ is the temperature (here set equal to 323°C, as in the experiments), and $V_a \approx 0.02$ nm³ is the volume per atom of pure Ge (any variation with the alloy composition is neglected). Mixing is modeled by the configurational entropy term, as for ideal alloys, neglecting enthalpic contributions (only expected to play a role for high Sn content). The chemical potential of each component at the surface is then obtained by definition:

$$\mu_i = \frac{\delta G}{\delta n_i} = \left(\frac{\delta G}{\delta \varphi} + \frac{1}{\epsilon}\tau(\hat{n})\frac{\partial \varphi}{\partial t}\right) + \frac{1-c}{\varphi}\frac{\delta G}{\delta c} =$$

$$= -\epsilon\nabla \cdot [\gamma\nabla\varphi] + \gamma\frac{1}{\epsilon}W'(\varphi) + \frac{1}{\epsilon}\tau(\hat{n})\frac{\partial \varphi}{\partial t} + \frac{kT}{V_a}\ln c_i$$

For numerical reasons[3,4], a stabilizing function $g(\varphi) = 10\varphi^2(1-\varphi)^2$ is put in front of $\mu_i$ when performing the simulations. Here we have included an additional term $\frac{1}{\epsilon}\tau(\hat{n})\frac{\partial\varphi}{\partial t}$ in the definition of the chemical potential $\mu_i$. This accounts for the kinetics of adatom redistribution on the surface, considering that each facet of the NW has a different incorporation time $\tau$, which quantifies the probability to incorporate an atom from the adatom phase into the crystal lattice. This quantity is modeled to be orientation dependent by means of the surface normal $\hat{n} = -\nabla\varphi/|\nabla\varphi|$, in order to introduce the dependency with the different crystal facets exposed by the NW, as shown in Fig. S2. It is worth noting that a single $\tau$ was used for the kinetic term, however in principle the two adatom species could be driven by different incorporation times.

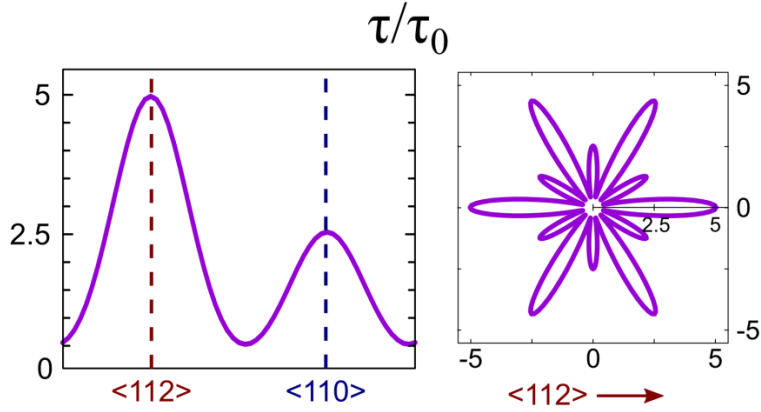

Figure S2. Plots of incorporation time $\tau(\hat{n})$ as a function of the crystal orientation, with particular respect to <112> and <110> orientations.

The coupled evolution equations for both profile shape and composition are then

$$\frac{\partial \varphi}{\partial t} = \Phi(\hat{n})|\nabla\varphi| + \nabla \cdot [M(\varphi)\nabla\mu_\varphi]$$

$$\mu_\varphi = \frac{\delta G}{\delta \varphi} + \frac{1}{\epsilon}\tau(\hat{n})\frac{\partial \varphi}{\partial t}$$

$$\frac{\partial(c\varphi)}{\partial t} = \Phi_{Sn}(\hat{n})|\nabla\varphi| + \nabla \cdot [cM(\varphi)\nabla\mu_\varphi] + \nabla \cdot M(\varphi)\nabla c$$

where $\Phi = \Phi_{Sn} + \Phi_{Ge}$ is the total growth rate and the last term in the third equation is the Fick diffusion within the interface region. The mobility function $M(\varphi) = (36/\epsilon)\varphi^2(1-\varphi)^2$ is non-zero only in the region of the $\varphi$ interface, i.e. at the NW surface, so that any material transfer causing a change in the shape or composition is only possible by surface diffusion. Bulk diffusion is suppressed so that the composition profile does not change in the region below the $\varphi$ interface. This model is equivalent to the one proposed in Ref. 5, but postulates equal mobilities for the two species and models the kinetic redistribution of adatoms.

Both Sn and Ge fluxes are set as functions of the surface orientation $\hat{n}$, exploiting the convenient formulation of Ref. 2. In particular, in order to have a flux composition of ≈10% Sn in the <110> directions and ≈15% Sn in the <112> ones, mimicking the experimental trend, a higher Sn flux is set on the {112} facets as reported in Fig. S3.

The flux distribution is compatible with the formation of a dodecagonal cross section in the kinetic crystal shape (KCS) approximation, where no diffusion is allowed on the surface, as shown in Fig. S3.

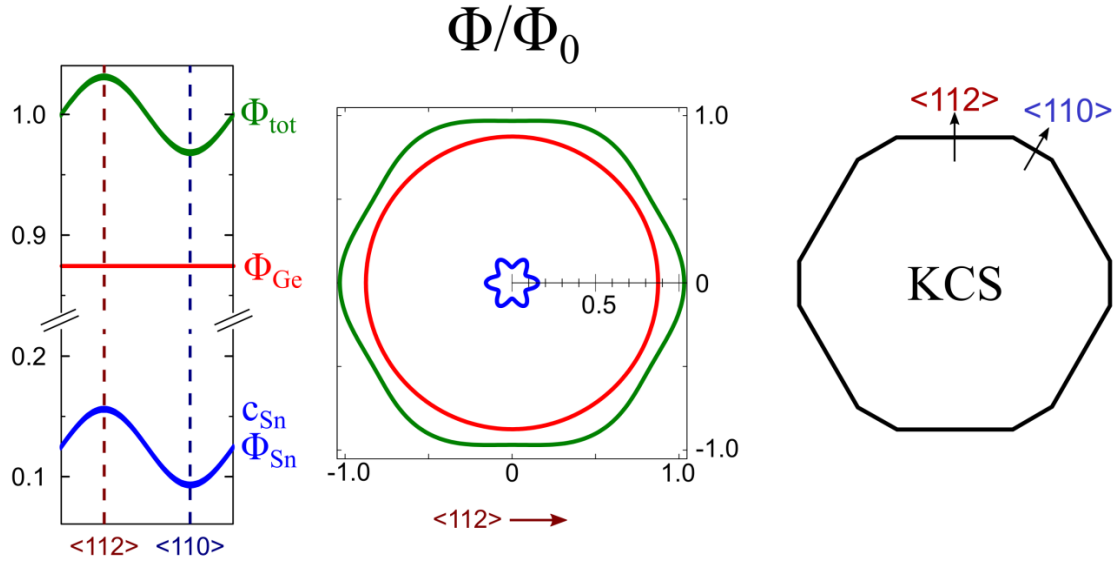

Figure S3. Plots of the total (Φ), Ge ($\Phi_{Ge}$) and Sn ($\Phi_{Sn}$) growth rates, with particular respect to <112> and <110> orientations, and the corresponding kinetic crystal shape (KCS).

Simulations are carried out by exploiting adaptive mesh refinement with maximum resolution (~0.7 nm) at the $\varphi$ interface and in correspondence of compositional variations. A semi-implicit integration scheme is implemented. Utilizing the system symmetry, only a quarter of the full NW section is considered. A dodecagonal Ge core of diameter 100 nm is considered as initial profile and the growth up to a 100 nm thick shell is simulated.

## S4. GeSn shell growth model – Composition profiles

In the Figure S4 we report the evolution of the composition profile at the shell surface (at $\varphi$ =0.5) as obtained by the simulations of Fig. 7 in the main manuscript.

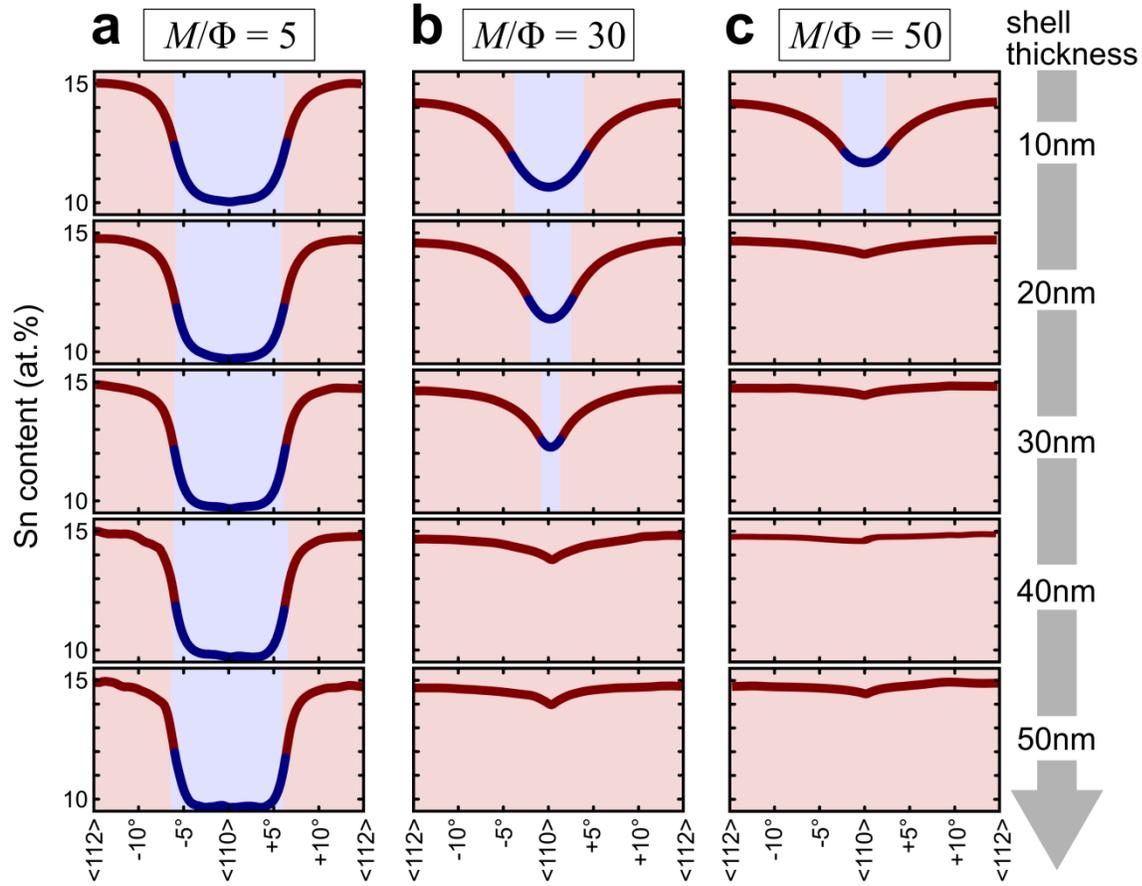

Figure S4. Composition profiles along the surface of the growing shell, between consecutive <112> directions, at the different deposition stages. The three cases (a,b,c) of Fig. 7 in the main manuscript are considered.

In the case (a), i.e. the simulation of Fig. 7a for smaller M/Φ ratio, the transition from the Sn-rich {112} region to the Sn-poor {110} one is rather abrupt since the diffusion length has a very short range. Still, for a thin shell (thickness <=20 nm), the Sn content in the <110> direction from the NW center is slightly higher than the nominal value from deposition (Fig. S3) because a tiny amount of Sn can still arrive there by diffusion from the borders with the {112} facets. But, as the thickness increases the length of the {110} front increases as well and Sn cannot reach its center anymore, so that the composition becomes the same of the deposition flux.

In case (b), i.e. the simulation of Fig. 7b, the diffusion of Sn extends over most of the {110} region so that there is a smoother transition in the composition from <112> to <110> direction. In particular, a certain amount of Sn can reach the middle of the {110} region so that the minimum of composition is at higher Sn content with respect to the nominal value from deposition. For the same reason, a lower Sn content is recognized even in the <112> direction.

As the shell thickness increases, the {110} segment shrinks in size and hence the Sn content at the center increases as also indicated by the radial plot in Fig. 7b. A variation in the composition profile is then observed through the whole {110} regions and also at the sides with the {112} ones, as perceivable also by the color transition in Fig. 7b of the main manuscript.

The Sn transfer toward <110> is even more evident for the case (c) where the Sn content at the center of the {110} region is enhanced since the early stages of the shell growth. Again, a decrease is also observed in the <112> direction due to mixing. The nominal composition is recovered as soon as the {110} segments disappear.